\documentclass{aa} 
\usepackage{graphics}
\begin{document} 

\thesaurus{4(11.13.2, 11.19.2, 11.09.1, 13.18.1, 02.16.2)} 
\title{ Regular magnetic fields in the dwarf irregular galaxy NGC~4449 } 

\author{K.T. Chy\.zy \inst{1}, R. Beck\inst{2}, S. Kohle\inst{3}, U. 
Klein\inst{3} and M. Urbanik\inst{1}}
\institute{Astronomical Observatory, Jagiellonian  University,  ul  Orla 
171, PL30-244 Krak\'ow, Poland 
\and 
Max-Planck-Institut f\"ur Radioastronomie, Auf dem H\"ugel  69,  D-53121 
Bonn, 
Germany 
\and
Radioastronomisches Institut der Universit\"at Bonn, Auf dem H\"ugel 71, 
D-53121 Bonn, Germany 
} 
\offprints{K. Chy\.zy (chris@oa.uj.edu.pl)} 
\mail{chris@oa.uj.edu.pl}
\date{Received date/ accepted date} 

\titlerunning{Regular magnetic fields in NGC~4449}
\authorrunning{K. Chy\.zy et al. }

\maketitle

\begin{abstract} 

We present a high-resolution VLA study of the total power and  polarized 
radio continuum emission at 8.46 and 4.86~GHz of  the  irregular  galaxy 
NGC~4449, known for its weak rotation and non-systematic gas motions. We 
found strong galaxy-scale regular magnetic fields, which  is  surprising 
because of a lack of ordered rotation required for  the  dynamo  action. 
The strength of the regular field reaches 8~$\mu$G and that of the total 
field 14~$\mu$G, comparable to that of the total magnetic field strength 
in radio-bright spirals. The magnetic vectors in  NGC~4449  form  radial 
``fans'' in the central region and fragments of a spiral pattern in  the 
galaxy's outskirts. These structures are associated with  large  regions 
of systematic Faraday rotation, implying genuine  galaxy-scale  magnetic 
fields rather than random ones compressed and stretched  by  gas  flows. 
The observed pattern of polarization B-vectors is similar to dynamo-type 
fields in normal spirals. Nonstandard, fast dynamo concepts are required 
to explain the observed field strengths, though it is unknown what  kind 
of magnetic field geometry can be produced  in  slowly  and  chaotically 
rotating objects. The so far neglected role of magnetic fields  for  the 
dynamics and star  formation  in  dwarf  irregulars  also  needs  to  be 
revised. 


\end{abstract} 

\section{Introduction} 

The generation of large-scale galactic magnetic fields from  small-scale 
field perturbations caused by turbulence (as postulated  by  the  dynamo 
concept) requires  a  preferred  sense  of  twisting  of  turbulent  gas 
motions, called the $\alpha$-effect (Wielebinski  \&  Krause  \cite{wielebinski}).  In 
normal spiral galaxies it is determined by Coriolis forces caused by the 
disk rotation giving rise to strong dynamo action and  to  the  observed 
spiral-like regular magnetic fields (Beck et  al.  \cite{beck96b}).  To  make  the 
dynamo process work, either the differential  rotational  shear  or  the 
galaxy's angular speed (in case of rigid rotation) must  exceed  certain 
threshold values (Ruzmaikin et al. \cite{ruzmaikin}). 

Dwarf irregulars are small, low-mass galaxies with a patchy distribution 
of star-forming regions. Though they exhibit a large variety of rotation 
curves (Hunter et al. \cite{hunter98a}) many of them show slow rotation  with  much 
less rotational shear than in normal spirals. Some dwarf irregulars show 
complex velocity fields with chaotic motions comparable in speed to  the 
overall  rotation.  Even  if  the  dynamo  could  still  work  in   such 
conditions, the generation time scales of the magnetic fields  estimated 
from classical dynamo theory would be very long and  strong  large-scale 
magnetic fields are not expected. Their  observational  detection  would 
mean that the dynamical role of global magnetic fields in  gas  dynamics 
and star formation in irregular galaxies has to be reconsidered. 

Signatures of a global magnetic field were already detected in the Large 
Magellanic Cloud (LMC, Klein et al. \cite{klein93}). However,  this  galaxy  still 
shows a significant degree of  differential  rotation  (Luks  \&  Rohlfs 
\cite{luks}) so that, like in normal spirals, the standard dynamo process could 
be at work. In this paper we  present  a  sensitive  radio  polarization 
study of the dwarf irregular galaxy NGC~4449 which  exhibits  only  weak 
signs of global rotation (cf. also Sabbadin et al. \cite{sabbadin84}, Hartmann et al. 
\cite{hartmann}). The radial velocities in NGC~4449 relative to  the  systemic  one 
reach  $\pm  20$  --  $30$~km/s.   However,   the   analysis   of   the 
high-resolution HI data cube (kindly supplied by Dr D. Hunter) does  not 
show the classical picture of a global rotation. Instead, NGC~4449 shows 
velocity jumps and gradients along both the major and  minor  axis  with 
centroids not coincident with the optical centre.  They  are  intermixed 
with chaotic velocity variations  with  an  amplitude  of  about  10  -- 
15~km/s. These very complex and chaotic kinematics, partly  due  to  the 
interactions with DDO125 (Hunter at al. \cite{hunter98b}) and possibly  also  to  a 
high star  formation  rate,  make  NGC~4449  an  interesting  target  to 
investigate the magnetic field structure under conditions very difficult 
for the classical galactic dynamo. The basic parameters of NGC~4449  are 
summarized in Table~\ref{tab-param}. 

\begin{table}[t]
\caption{Basic properties of NGC 4449} 
\label{tab-param}
\begin{flushleft}
\begin{tabular}{lll}
\hline\noalign{\smallskip}
Type & IBm\\
R.A. (1950) & $12^h 25^m 44.8^s$ & LEDA database\\
Decl. (1950) & $44\degr 22\arcmin 08\arcsec$ & LEDA database\\
R.A. (2000) & $12^h 28^m 11.3^s$ & LEDA database\\
Decl. (2000) & $44\degr 05\arcmin 30\arcsec$ & LEDA database\\
Opt. extent$^*$) & $5.8\arcmin$ x $4.5\arcmin$ & LEDA database\\
Position angle & $45\degr$  & LEDA database\\
Inclination & $43\degr$ & Tully (\cite{tully})\\
Distance & 3.7 Mpc & Bajaja et al. (\cite{bajaja})\\
&$1\arcmin$ corresponds to & 1.08~kpc \\
Abs. B-magn. & $-18.5$ & Schmidt \& Boller (\cite{schmidt})\\
$M_{tot}$ & $7 \cdot 10^{10}M_{\sun}$ & Bajaja et al. (\cite{bajaja})\\
$M_{HI}$ & $2.3 \cdot 10^{9}M_{\sun}$ & Bajaja et al. (\cite{bajaja})\\
\noalign{\smallskip}
\hline
\end{tabular}
\end{flushleft}
$^*$) isophotal diameter at 25$^{m}/(\sq\arcsec)$
\end{table}
A low-resolution detection of polarized emission  (Klein  at  al. \cite{klein96}) 
showed that the magnetic field in NGC~4449 is running across its  bright 
star-forming body, very different from that in normal galaxies. In  this 
work we present a total power and polarization study of this galaxy with 
a resolution and sensitivity several times better than that in the  work 
of Klein et al. (\cite{klein96}). The use of two frequencies (8.46  and  4.86~GHz) 
enables us to determine the distribution of Faraday  rotation  over  the 
disk of NGC~4449, allowing  to  discriminate  between  the  galaxy-scale 
uniform fields and those passively stretched and compressed in  the  gas 
flows powered by huge star-forming regions. 

\section{Observations and data reduction} 

\begin{figure*}
\resizebox{13cm}{!}{\includegraphics{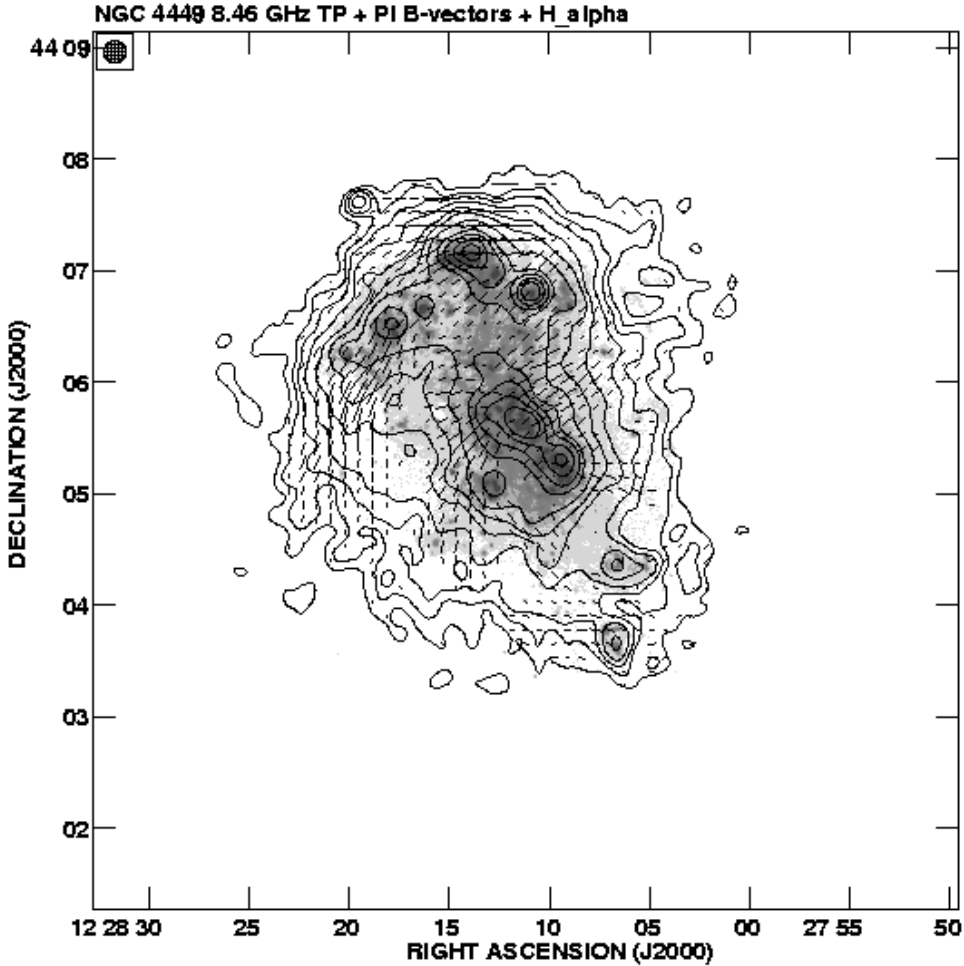}}
\parbox[b]{45mm}{
\caption{ 
The total power contour  map  of  NGC~4449  at  8.46~GHz  with  observed 
B-vectors of polarized intensity (taken as perpendicular  to  E-vectors) 
superimposed onto the H$\alpha$ image of Bomans et al. (\cite{bomans}), digitally 
enhanced  to  show  the  low-brightness  details  of  the  ionized   gas 
distribution. The resolution of the radio map is 12\arcsec. The  contour 
levels are  (3,  5,  7,  10,  15,  20,  30,  60,  90,  140,  200)$\times 
13~\mu$Jy/b.a., the r.m.s. noise level of the total power  map.  Vectors 
of 10$\arcsec$ correspond to 50~$\mu$Jy/b.a. 
} 
\label{8tp} }
\end{figure*} 

The maps of  total  power  and  linearly  polarized  radio  emission  of 
NGC~4449 at 8.46~GHz and 4.86~GHz were obtained  using  the  Very  Large 
Array  (VLA)  of  the  National  Radio  Astronomy   Observatory   (NRAO) 
\footnote{The National Radio Astronomy Observatory is a facility of  the 
National Science Foundation  operated  under  cooperative  agreement  by 
Associated Universities, Inc.}. To attain highest sensitivity to  smooth 
extended structures the most compact (D)  configuration  was  used.  The 
observations were carried out on 30 August and 1  --  2  September  1996 
using 27 antennas at two independent IFs, each with a  bandwidth  of  50 
MHz, separated by 50 MHz, with 11$^h$ integration time on  NGC  4449  at 
6.2~cm and 16$^h$ at 3.5~cm. The data were reduced  using  the  standard 
AIPS software package. The flux density scale and the position angle  of 
polarization was calibrated  by  observing  the  point  source  3C  286. 
Instrumental polarization was corrected by observing 1216+487, which was 
also used for gain and phase  calibration.  The  calibrated  and  edited 
visibility data were cleaned and self-calibrated (in phase  only)  using 
the AIPS package, yielding maps of Stokes parameters I, Q and U. 

These IUQ data were combined  with  Effelsberg  measurements  using  the 
program EFFMERG, a version of the SDE task IMERG (Cornwell et al. \cite{cornwell}) 
modified  by  P.~Hoernes  (see  Beck  \&  Hoernes \cite{beck96a}).  This  program 
deconvolves both clean maps with their  beams,  and  Fourier  transforms 
them back into the UV plane. Then a combination of Effelsberg  data  for 
small spacings and VLA data for  large  spacings  is  performed  with  a 
linear interpolation in  the  overlap  domain.  The  combined  data  are 
transformed back into the image plane with the synthesized VLA beam.  To 
avoid ring-like distortions around strong unresolved sources, introduced 
by the combination technique, the bright unresolved sources  were  first 
subtracted from the maps and added again after the combination (see Beck 
et al. \cite{beck97}). For our 8.46~GHz map we used the single dish data at 10.55 
GHz from Klein et al. (\cite{klein96}), scaled to  our  frequency.  We  used  mean 
spectral index of $-0.7$ (S$_{\nu}\propto \nu^{\alpha}$).  For  4.86~GHz 
we performed separate  observations  with  the  100-m  Effelsberg  radio 
telescope at 4.85~GHz. 

The Q and U maps were combined to get maps  of  the  linearly  polarized 
emission (corrected for the positive  zero  level  offset)  and  of  the 
position angle of  polarization  E-vectors.  The  final  maps  have  the 
synthesized beam of $12\arcsec$ at 8.46~GHz and $19\arcsec$ at 4.86~GHz. 
To obtain the map of Faraday rotation the data at both frequencies  were 
convolved to a common beam of $19\arcsec$. 

\section{Results} 

\subsection{Total power and polarized emission at 8.46~GHz } 

The total power map of NGC~4449 at 8.46~GHz with apparent  B-vectors  of 
polarized intensity is shown in Fig.~\ref{8tp}.  As  no  correction  for 
Faraday rotation was applied, the orientation of observed B-vectors  may 
differ from the magnetic field directions by some $3\degr$  --  $6\degr$ 
on average in the disk, the maximum difference reaching $\simeq 10\degr$ 
in small regions of high Faraday rotation measures (see Sect. 3.3).  The 
map shows details of the radio structure in the inner  disk.  The  total 
power emission shows strong peaks at the position of bright star-forming 
regions. In addition to  that  diffuse  radio  emission  away  from  the 
optically bright star-forming body has been detected as well. This radio 
envelope  extends  along  the  galaxy's  minor  axis  up  to  $2\arcmin$ 
(corresponding to $\simeq 2.2$~kpc) from the main plane. The  extent  of 
the radio envelope at 8.46~GHz is larger than that of the faint  diffuse 
H$\alpha$ emission (see Fig.~\ref{8tp}). In the southern disk the  radio 
emission has an extension towards a  nebulous  object  at  RA$_{2000}  = 
12^{h}  28^{m}  06\fs7$,  Dec$_{2000}  =  44\degr  03\arcmin  39\arcsec$ 
(probably a supernova remnant), forming a faint peak at its position. 

The  contour  map  of  polarized  brightness  with  apparent   B-vectors 
proportional to the polarization degree is shown in Fig.~\ref{8pi}.  The 
extended radio emission is substantially polarized (locally up to 50\%), 
with extended ($\ge 1$~kpc) domains of highly  aligned  B-vectors.  The 
magnetic field structure in the inner disk looks unusual at first glance 
(Figs.~\ref{8tp} and \ref{8pi}). The projected magnetic vectors  in  NGC~4449  show  two 
distinct kinds of structure. From the bright central star-forming region 
they are directed radially outwards, on each side  forming  a  polarized 
``fan''. The B-vectors are parallel to the H$\alpha$ filaments discussed 
in detail by Sabbadin \& Bianchini (\cite{sabbadin79}) and by Bomans et 
al. (\cite{bomans}). 
In the galaxy's outskirts, the magnetic vectors run  along  a  polarized 
ridge encircling the galaxy on the northern, north-eastern  and  eastern 
side. Between this  structure  and  the  eastern  ``fan''  an  elongated 
unpolarized ``valley'' is due to a geometrical superposition of mutually 
perpendicular  polarization  directions  in  the  ``fan''  and  in   the 
polarized ridge. 

\subsection{Total power and polarization at 4.86~GHz} 

\begin{figure*}
\resizebox{13cm}{!}{\includegraphics{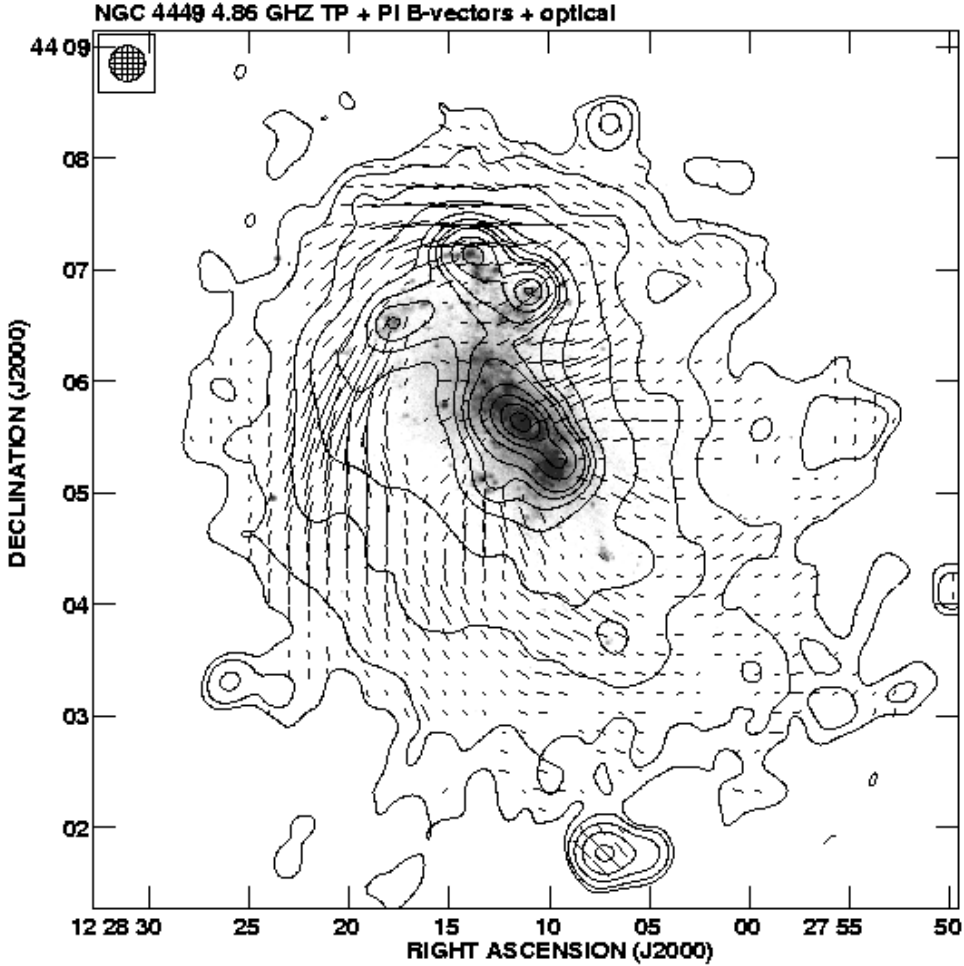}}
\parbox[b]{45mm}{
\caption{ 
The contour map of  the  total  power  of  NGC~4449  at  4.86~GHz   with 
B-vectors of the polarized intensity superimposed onto an optical  image 
obtained by one of us (SK) at the Hoher List Observatory. The resolution 
is 19$\arcsec$. The contour levels are (3, 5, 10, 20, 40, 60,  80,  100, 
150, 200, 250, 300)$\times 26~\mu$Jy/b.a., the r.m.s. noise in the total 
power map. Vectors of  10$\arcsec$  correspond  to  50~$\mu$Jy/b.a..  No 
correction for Faraday rotation was applied (see Sect. 3.2) 
} 
\label{6tp} }
\end{figure*}

Our maps at 4.86~GHz have a considerably worse resolution than those  at 
8.46~GHz, and the orientations  of  the  B-vectors  may  be  subject  to 
stronger Faraday rotation (on average about $10\degr$ but locally up  to 
$\simeq 30^{o}$, see Fig.~\ref{rmm}). However, due to a  higher  signal-to-noise 
ratio at  4.86~GHz  the  radio  emission  is  traced  much  further  out 
(Fig.~\ref{6tp}). At this frequency we can trace the radio  envelope  in 
the sky plane out to $3\farcm 2$ (3.5~kpc) from the galaxy's major axis. 
The nonthermal emission thus extends into the halo beyond one  isophotal 
(at the level of 25$^{m}/(\sq\arcsec)$) major axis radius,  which  is  a 
rare (though not exceptional) phenomenon  among  spiral  galaxies  (e.g. 
Hummel et al. \cite{hummel}). 

\begin{figure}
\resizebox{\hsize}{!}{\includegraphics{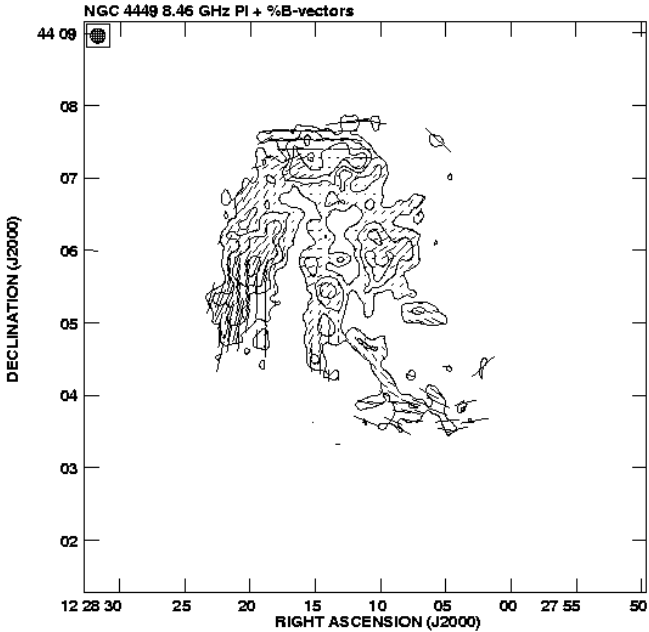}}
\caption{
The contours  of  polarized  intensity  of  NGC~4449  at  8.46~GHz  with 
superimposed B-vectors (perpendicular to the observed E-vectors) of  the 
polarization  degree.  The  contour  levels  are  (2,  4,  6,  9)$\times 
9~\mu$Jy/b.a., the r.m.s. noise of the polarized intensity map. Vectors 
of 10$\arcsec$ correspond to 20\%}
\label{8pi}
\end{figure}

\begin{figure*} 
\resizebox{13cm}{!}{\includegraphics{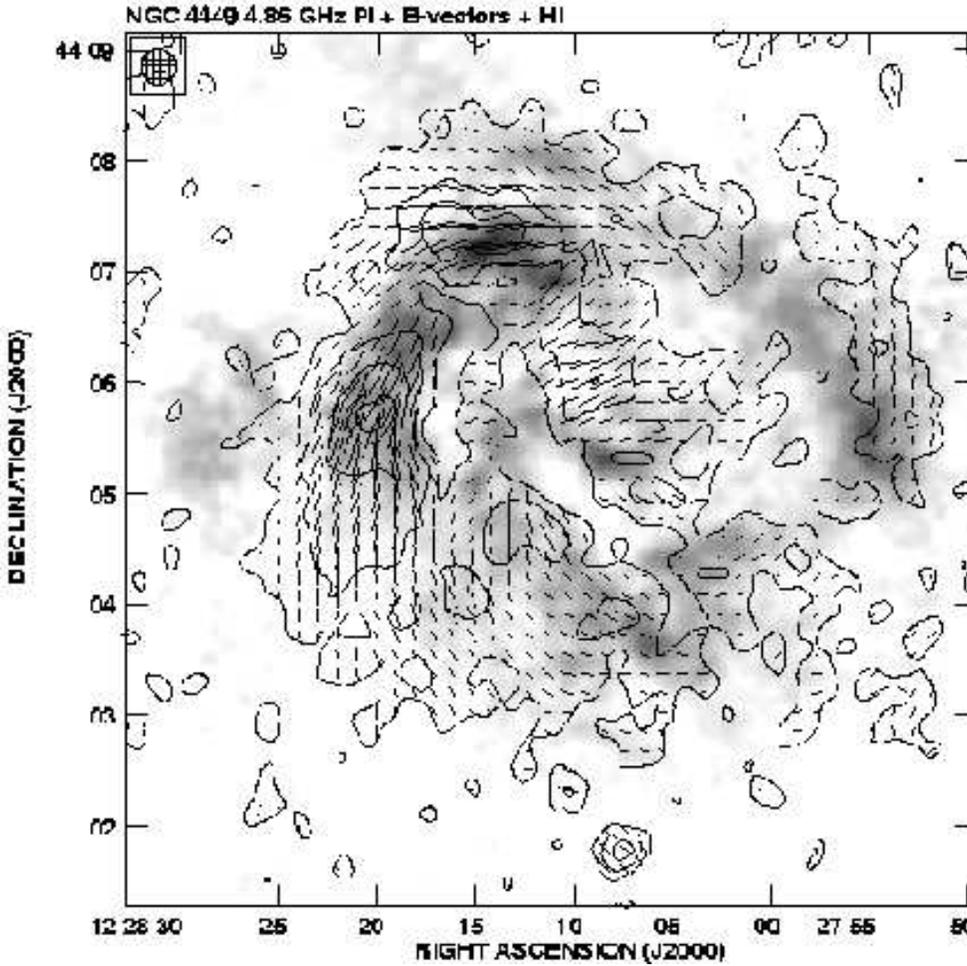}}
\parbox[b]{45mm}{
\caption{ 
Contours and  B-vectors  of  the polarized  intensity  of  NGC~4449   at 
4.86~GHz with a resolution 19$\arcsec$ superimposed onto a  colour  plot 
of the HI column density made from the data kindly supplied by Dr.  D.A. 
Hunter from the Lovell Observatory. The contour levels of the  polarized 
intensity are (3, 10, 20, 30,  35)$\times  5.4~\mu$Jy/b.a.,  the  r.m.s. 
noise level in the polarized intensity map 
} 
\label{6pi} }
\end{figure*}

The map at 4.86~GHz again shows  the  polarized  ``fans'',  however  the 
eastern one is less conspicuous at  this  frequency  than  at  8.46~GHz, 
which suggests stronger  Faraday  depolarization  in  this  region.  The 
polarized ridge in the  northwestern  portion  of  the  galaxy,  already 
visible in Fig.~\ref{8pi}, turns out to be part of  a  larger  polarized 
ring surrounding the galaxy from the northeast through north,  east  and 
south down to the southwest, with a well-organized, coherent pattern  of 
magnetic  vectors  (Fig.~\ref{6pi}).  Another  weak  fragment   of   the 
polarized ring is visible west of the centre. The  ring  coincides  well 
with  a  similar  feature  visible   in   HI   (Hunter,   priv.   comm., 
Fig.~\ref{6pi}), with one of  the  brightest  polarization  peaks  lying 
close to the densest neutral gas clump. Along the ring the  polarization 
B-vectors are not exactly tangential to the azimuthal directions  or  to 
the HI shell. They deviate systematically from the azimuthal  directions 
by some $20\degr$ -- $40\degr$. A detailed discussion  of  the  magnetic 
field directions is presented in Sect.~4. 

Due to the higher sensitivity at 4.86~GHz our map shows  very  well  the 
unpolarized ``valley'' not only at the interface of the eastern  ``fan'' 
and the ridge but also a similar feature in the NW disk. In  both  cases 
they  result  from  a  geometrical  superposition  of   magnetic   field 
directions in the ``fans''  and  in  the  polarized  ring,  seen  almost 
perpendicular to each other when projected to the sky plane. 

\subsection{Faraday rotation }

The distribution of Faraday rotation measures between 8.46 and  4.86~GHz 
is shown in Fig.~\ref{rmm}.  The  northern  and  eastern  parts  of  the 
polarized ring, as well as the ``magnetic  fan''  east  of  the  central 
star-forming complex show coherently positive Faraday rotation  measures 
(RM) over areas with sizes of about $1.5\arcmin$, with a mean  value  of 
about +50 rad/m$^{2}$. The  values  of  RM  are  rising  locally  up  to 
+200~rad/m$^{2}$. The western ``fan''  and  the  southern  part  of  the 
polarized ring are dominated  by  negative  RMs,  on  average  of  about 
$-50$~rad/m$^{2}$ but  also  reaching  $-150$~rad/m$^{2}$  locally.  The 
errors in these regions vary from $\pm 10$ to $\pm  20$  rad/m$^{2}$  in 
regions of low RMs, exceeding $\pm 50$ rad/m$^{2}$ in  regions  of  high 
rotation measures. However, though in individual points the values of RM 
do not generally exceed the errors by more than 2 -- 2.5$\sigma$  r.m.s. 
errors, coherent areas of the same sign of  RM  extend  over  many  beam 
sizes. The statistical significance  of  our  determinations  of  RM  is 
discussed in detail in Sect.~4. 

\section{Discussion} 

\subsection{Distribution of thermal emission}
In order to determine  the  equipartition  magnetic  field  strength  in 
selected regions of NGC~4449 we need to  estimate  the  distribution  of 
thermal emission in the galaxy. The spectral index computed between  the 
maps of NGC~4449 at 8.46~GHz and  4.86~GHz  changes  from  about  $-0.3$ 
(S$_{\nu}\propto \nu^{\alpha}$)  in  strongly  star-forming  regions  to 
$-1.1$ locally in the outer southern region. Somewhat smaller variations 
are found by Klein et al. (\cite{klein96}), probably because  of  the  much  lower 
resolution used by these authors. Although Klein  et  al.  (\cite{klein96})  found 
variations of the nonthermal spectral index $\alpha_{nt}$ between $-0.5$ 
in the central star-forming region to $-0.8$ in the eastern part of  the 
halo, for our purpose it was sufficient to assume $\alpha_{nt}$ constant 
over the whole galaxy. Possible uncertainties  due  to  this  assumption 
were included in the errors. To determine the nonthermal spectral  index 
we compared the radial distribution of the thermal  brightness  S$_{th}$ 
at 8.46~GHz and that in the H$\alpha$ line (S$_{H\alpha}$), convolved to 
the  beam  of  $19\arcsec$.  We  found  that  they  are  identical   for 
$\alpha_{nt}$ = $-0.9$. This value differs  only  by  about  1.5$\sigma$ 
r.m.s. from the value obtained by Klein et al.  (\cite{klein96})  from  the  radio 
spectrum, but agrees better with their estimate based  on  thermal  flux 
obtained from the H$\alpha$ emission. 

\begin{figure}[t] 
\resizebox{\hsize}{!}{\includegraphics{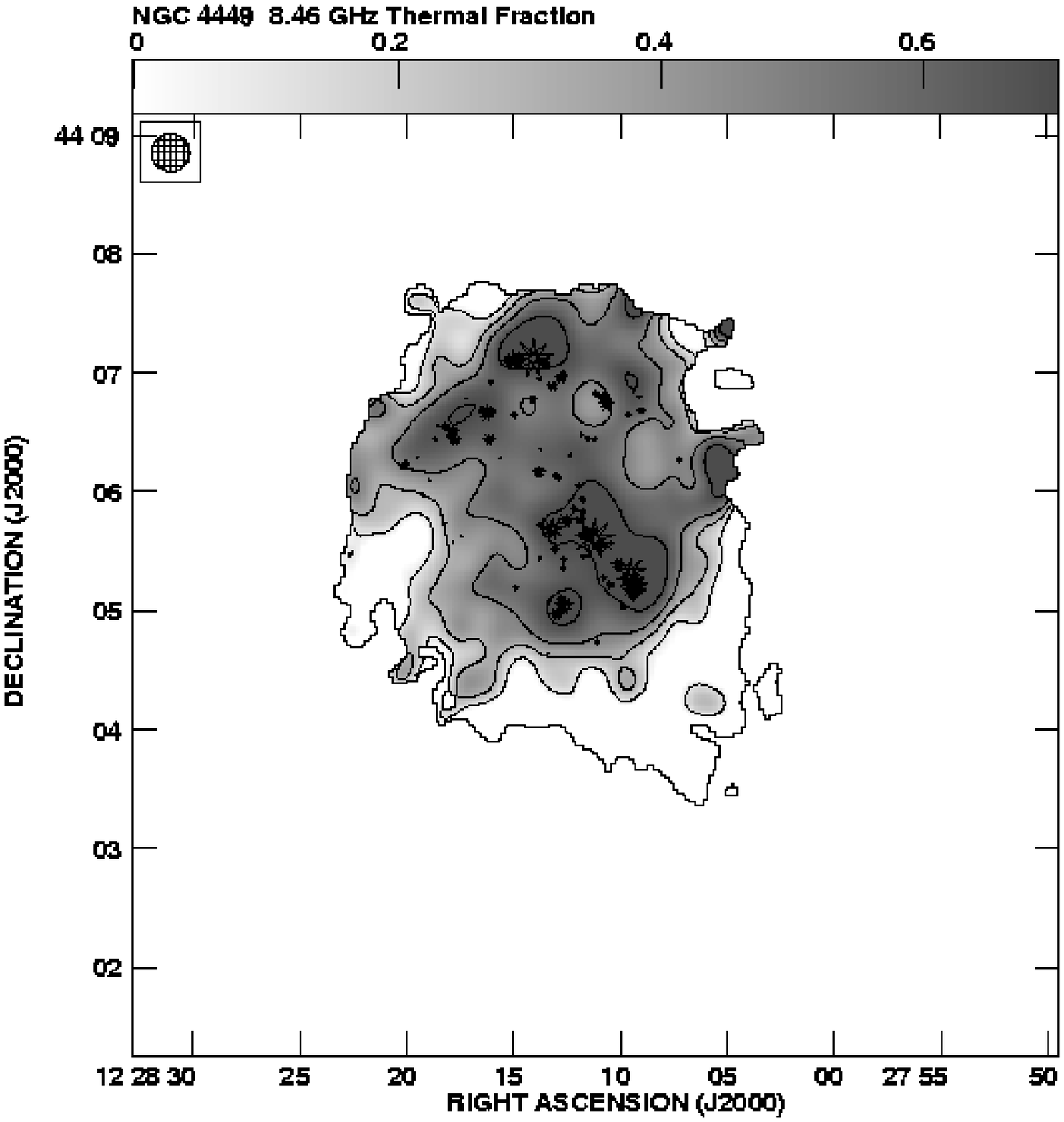}}
\caption{ 
The distribution of  thermal  fraction  in  NGC~4449  derived  from  the 
distribution of the spectral index between 8.46~GHz  and  4.86~GHz.  The 
data at both frequencies are convolved to a common beam of  $19\arcsec$. 
The mean nonthermal spectral index  of  $-0.9$  has  been  adopted.  The 
contour levels are: 0.1, 0.3, 0.5, 0.7. Symbols mark  the  positions  of 
the brightest HII regions 
} 
\label{therm} 
\end{figure}

The distribution of thermal fraction f$_{th}$ at  8.46~GHz  in  NGC~4449 
(Fig.~\ref{therm})  shows  clear  peaks  at  the  positions  of   bright 
star-forming complexes, f$_{th}$ reaches 80\% there.  After  subtraction 
of the thermal emission these regions are  still  considerably  brighter 
than the diffuse emission from the surroundings by some 40\%. Away  from 
bright star-forming complexes the emission is  largely  nonthermal,  the 
free-free emission amounts to not more than 10\%. 

\begin{figure}[t] 
\resizebox{\hsize}{!}{\includegraphics{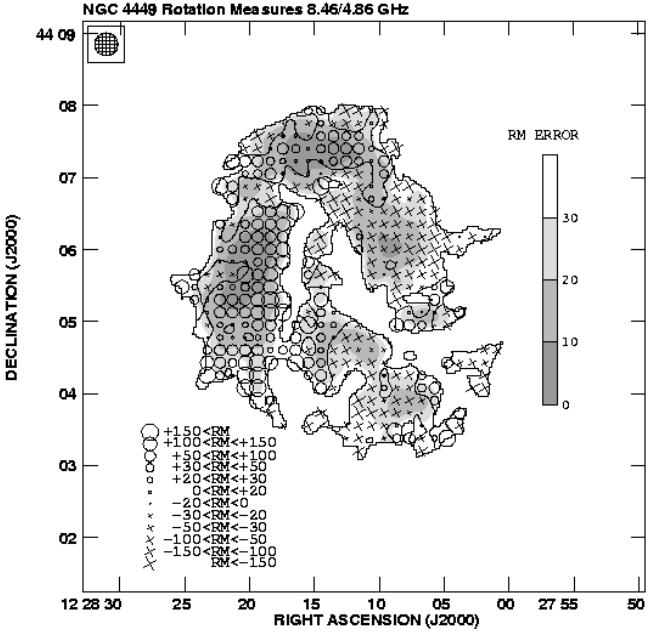}}
\caption{ 
The distribution of the Faraday rotation measures (RM) in  the  disk  of 
NGC~4449, computed between 8.44 and 4.86~GHz. All data were convolved to 
a common beam of 19\arcsec. Positive  and  negative  values  of  RM  are 
marked by circles and crosses, respectively. The symbol  sizes  indicate 
the absolute value of RM as indicated in the figure legend. The  contour 
line divides the regions with positive and negative values  of  RM.  The 
underlying greyscale plot shows the RM errors 
}
\label{rmm} 
\end{figure} 

As an additional test of our assumption of $\alpha_{nt}$  constant  over 
the galaxy's body we analyzed  the  point-to-point  correlation  between 
maps of the radio thermal flux at 8.46~GHz and  that  in  the  H$\alpha$ 
line convolved to $19\arcsec$. We checked that,  using  the  Monte-Carlo 
simulations of two-dimensional arrays of  points  convolved  to  various 
beams, values in map points separated by 1.2 times  the  beam  size  are 
correlated only by some 10 -- 12\%  and  are  for  our  purposes  almost 
independent. Therefore we used points separated by $22\fs 8$.  In  order 
to eliminate an artificial correlation caused by the radial decrease  of 
all quantities, each map was divided by an axisymmetric  model  obtained 
by integrating the map in elliptical rings with the position  angle  and 
inclination taken from Tab.~\ref{tab-param}. A correlation slope significantly  larger 
than 1 would mean that we have overestimated the thermal radio  emission 
in strongly star-forming regions while in fact they  have  much  flatter 
nonthermal spectra. Using the orthogonal regression we obtained  S$_{th} 
\propto  $  S$_{H\alpha}^{0.96\pm  0.09}$,  thus  close  to   a   linear 
relationship, though few regions  deviate  strongly  from  the  best-fit 
line.  This  means  that  $\alpha_{nt}$  shows   some   place-to   place 
variations, but our assumption of  $\alpha_{nt}$  =  constant  does  not 
introduce large, systematic errors in determining the thermal  fraction. 
A detailed multi-dimensional analysis of both radio emission  components 
involving the H$\alpha$, CO, HI and X-ray data will be the subject of  a 
separate study. 

\subsection{Magnetic field strengths} 

To  determine  the  magnetic  field  strengths  in  NGC~4449  from   the 
synchrotron emission we assumed  the  equipartition  conditions  between 
magnetic fields and cosmic rays to be valid everywhere  in  the  galaxy. 
Furthermore we adopted a proton-to-electron ratio of energy densities of 
100 and a lower energy cutoff of cosmic ray electrons  of  300  MeV.  We 
assumed a face-on thickness of the nonthermal disk of  2~kpc,  resulting 
from a typical scale height of galactic radio disks of 1~kpc (Hummel  et 
al. \cite{hummel}), determined by the propagation range of cosmic ray  electrons. 
With the inclination of  NGC~4449  from  Tab.~\ref{tab-param}  this  implies  a  mean 
pathlength through the  galaxy  of  2.8~kpc.  The  errors  of  estimated 
magnetic field strengths include an uncertainty of these quantities of a 
factor two. The thermal fractions were taken from results  described  in 
Sect. 4.1.

\begin{table}[t]
\caption{Magnetic fields in NGC 4449} 
\label{tab-field}
\begin{flushleft}
\begin{tabular}{lccrr}
\hline\noalign{\smallskip}
Region & Position & & total &regular \\
 & & & field & field \\
 & RA$_{2000}$ & Dec$_{2000}$ & $\mu$G & $\mu$G\\
\hline\noalign{\smallskip}
 Average for & NGC~4449 & & $ 12\pm 4$ & $4\pm 1$ \\
NW ridge & $12^h 28^m 14.5^s$ & $44\degr 07\arcmin 15\arcsec$
& $14\pm 4$ & $7\pm 3$ \\
Eastern ridge & $12^h 28^m 20.5^s$ & $44\degr 05\arcmin 47\arcsec$ 
& $13\pm 4$ & $8\pm 3$ \\
Western fan & $12^h 28^m 09.1^s$ & $44\degr 05\arcmin 51\arcsec$ 
& $14\pm 4$ & $7\pm 2$ \\
Eastern fan & $12^h 28^m 14.2^s$ & $44\degr 05\arcmin 29\arcsec$ 
& $14\pm 5$ & $6\pm 2$ \\
\noalign{\smallskip}
\hline
\end{tabular}
\end{flushleft}
\end{table}

Under these assumptions we determined the mean magnetic  field  strength 
for  the  whole  galaxy  and  for  selected  regions;  the  results  are 
summarized in Tab.~\ref{tab-field}. Regular magnetic fields derived from the polarized 
intensity were found to reach locally up to $7\pm 2~\mu$G in the western 
magnetic ``fan'' and about $8\pm 3~\mu$G in the radio-bright part of the 
polarized ring. The total magnetic field in  these  regions,  determined 
from the total power emission reaches $14\pm 4~\mu$G, comparable to that 
in the radio-brightest spiral  galaxies  (Beck  et  al.  \cite{beck96b}).  A  slow 
rotation of NGC~4449 accompanied by chaotic gas motions apparently  does 
not exclude the existence of strong, regular magnetic fields. 

\subsection{Magnetic field structure }

\subsubsection{Magnetic field coherence}

The presence of polarized emission alone does  not  provide  a  definite 
proof  for  dynamo-generated,  spatially   coherent   magnetic   fields. 
Substantial polarization may be also produced  by  random  fields,  made 
anisotropic by  squeezing  or  stretching,  e.g.  by  stellar  winds  or 
large-scale shocks from multiple  supernova  events,  however,  frequent 
field reversals along the line of  sight  would  completely  cancel  the 
Faraday rotation. Non-zero rotation measures imply the  magnetic  fields 
in the observed  galaxy  coherent  over  scales  much  larger  than  the 
telescope beam. 

Although the values of RM in individual points of our  Faraday  rotation 
map (Fig.~\ref{rmm}) do not exceed the errors by much, we note that they 
deviate coherently from zero, forming large domains of constant RM  sign 
(both positive and negative). These regions with mean RM of $\simeq  \pm 
50$~rad/m$^{2}$ are up to 20 times larger than the telescope beam  area. 
A correction  for  the  foreground  rotation  of  $-35$~rad/m$^{2}$  was 
estimated from background sources present in our map  and  checked  with 
the galactic RM map by  Simard-Normandin  \&  Kronberg  (\cite{simard}).  At  the 
galactic latitude of NGC~4449 of $72\degr$ the existence  of  foreground 
rotation structures changing sign over angular scales of  $2\arcmin$  -- 
$4\arcmin$  with  an  amplitude  of  100~rad/m$^{2}$,  correlated   with 
particular features in the galaxy's polarized  intensity,  is  unlikely. 
Thus, the observed Faraday effects  almost  certainly  originate  inside 
NGC~4449. 

To check quantitatively the coherence of the non-zero  Faraday  rotation 
we computed values of RM and its error $\sigma_{RM}$ in a grid of points 
separated by $22\fs 8$ (1.2 times the beam size). In case of a  lack  of 
systematic  Faraday  rotation  such  points  would  show   only   little 
correlation (see Sect. 4.1) and the variable defined as RM/$\sigma_{RM}$ 
is expected to fluctuate randomly from point to point with a  zero  mean 
and unity variance. However, we found that its mean value deviates  from 
zero in the eastern polarized ridge by more than  $4.1\sigma_{mean}$  as 
well  as  in  the  western  ``fan''  by  more  than  $4.3\sigma_{mean}$, 
$\sigma_{mean}$ being the r.m.s. error of mean RM  in  a  given  region. 
This implies that the probability  of  creating  at  random  such  large 
non-zero RM domains is less  than  $10^{-5}$.  In  the  eastern,  weaker 
``fan'',  the  deviation  amounts  to   only   $1.6\sigma_{mean}$   (the 
probability of a random occurrence of non-zero RM of 10\%), because of a 
worse signal-to-noise ratio. The results were found to be independent of 
the assumed foreground rotation. Thus we conclude that NGC~4449 contains 
genuine unidirectional fields,  rather  than  stretched  and  compressed 
random  magnetic  field.  The   latter   one   would   have   different 
sky-projected components yielding a substantial polarization  while  the 
line-of-sight component would frequently change sign which would  cancel 
any systematic Faraday rotation. We note that the growth of galaxy-scale 
coherent, unidirectional fields lies at the foundations  of  the  dynamo 
process. 

\subsubsection{Magnetic field geometry}

\begin{figure*} 
\resizebox{13cm}{!}{\includegraphics{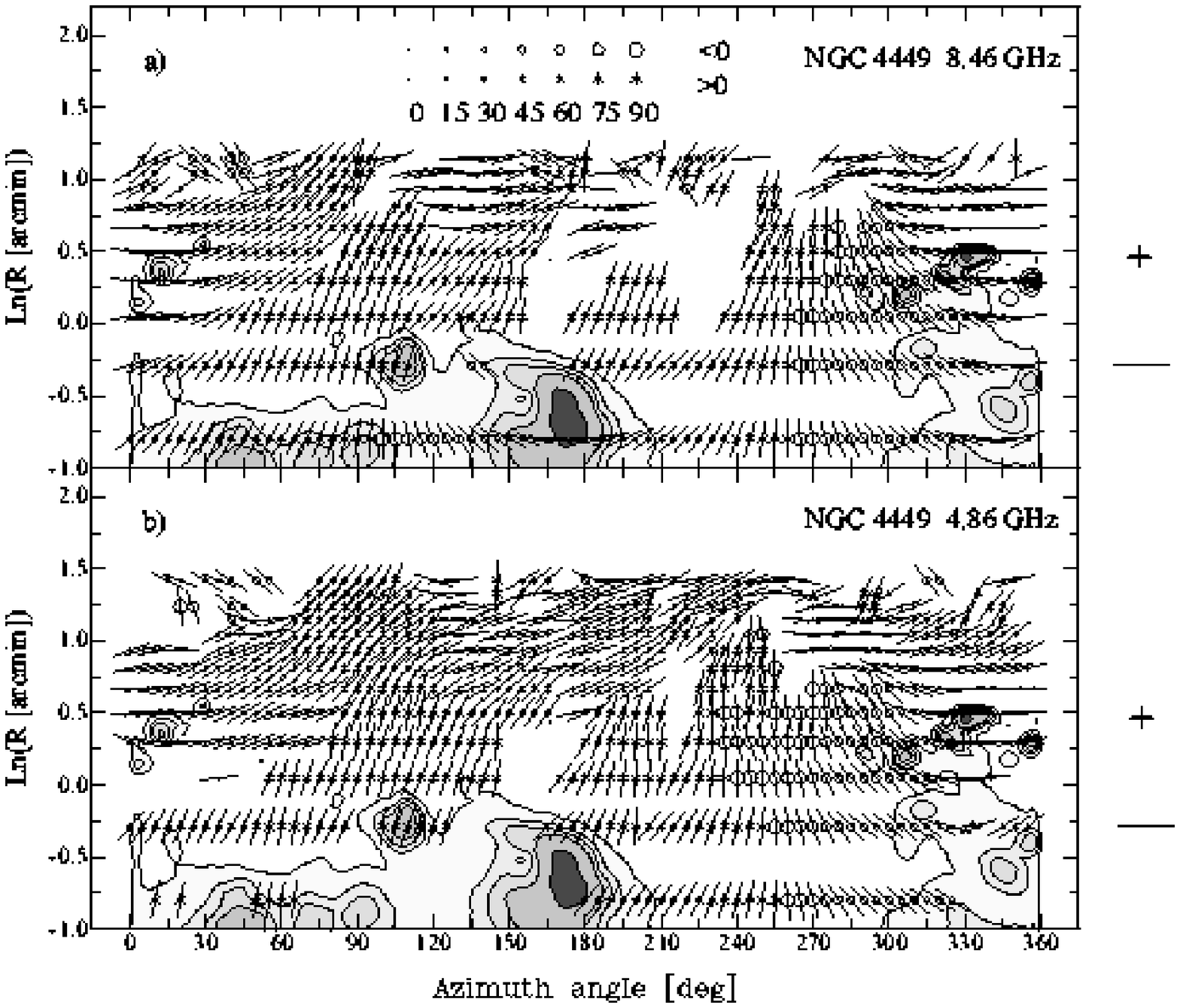}}
\parbox[b]{45mm}{
\caption{
The distribution of magnetic pitch angles in the  disk  of  NGC~4449  at 
8.46~GHz (a) and 4.86~GHz (b) as a function of azimuthal  angle  in  the 
disk and ln(R), R being the galactocentric radius in  arcmin.  The  data 
were corrected for the galaxy inclination taken from the  LEDA  database 
and  refer  to  the  galaxy's  main  plane.  The  azimuthal  angle  runs 
counterclockwise from the NE end of the major axis. The  greyscale  plot 
shows the distribution of the H$\alpha$ emission (Bomans  et  al. \cite{bomans}) 
convolved to a  Gaussian  beam  of  $19\arcsec$.  The  size  of  symbols 
superimposed on the polarization B-vectors is proportional to the  value 
of the magnetic pitch angle (see the legend in the Figure) 
}
\label{pitch} }
\end{figure*}

Fig.~\ref{pitch} a and b presents the  distribution  of  magnetic  field 
orientations in the azimuth-ln(R) frame (R  being  the  radial  distance 
form galaxy's optical centre), in which the logarithmic  spiral  appears 
as a set of straight lines inclined by  the  spiral's  pitch  angle.  At 
8.46~GHz (little Faraday rotation) we clearly see a combination  of  the 
radial field in the inner region out to ln(R) of 0.5 -- 0.6 and  a  more 
azimuthal one at larger radii. However, at this frequency the picture in 
the  outer  galaxy  regions  becomes  rather  noisy.  A  comparison   of 
Figs.~\ref{pitch} a and b shows that  Faraday  rotation  does  not  much 
change the global field picture in the inner region  where  the  ionized 
gas density is highest and Faraday effects strongest. Thus, the 4.86~GHz 
data alone can be safely used in the galaxy outskirts. 

Fig.~\ref{pitch}b shows a very well ordered field in the polarized  ring 
with the magnetic pitch angle $\psi$ keeping a constant sign  over  most 
of azimuthal angles (except a low signal-to-noise region at azimuths  of 
$0\degr$ -- $60\degr$ and ln(R) $\ge 1$). The value of $\psi$ is $\simeq 
40\degr$ on average, with local  variations.  It  resembles  a  somewhat 
distorted magnetic spiral with a  substantial  radial  component.  This, 
like in rapidly rotating spiral galaxies, may signify dynamo-type fields 
(Urbanik et al. \cite{urbanik}), while the  random  field  pushed  away  from  the 
galaxy and squeezed by  an  expanding  gaseous  shell  would  yield  the 
observed B-vectors parallel to the shell. Nevertheless, the pitch angles 
show some place-to-place changes, possibly due to processes  like  local 
outflows or compressions. The strongest distortion of the spiral  -  the 
region of nearly pure toroidal magnetic field at azimuthal  angles  $\ge 
270\degr$, ln(R) $\ge 0.5$ coincides with the densest  HI  clump  and  a 
region of star formation. The analysis of recent CO data (Kohle  et  al. 
in preparation) suggests strong gas compression possibly due to external 
interactions. We note also an opposite sign of Faraday rotation at  both 
ends of the major axis,  which  is  typical  for  axisymmetric  magnetic 
fields.

The radial magnetic ``fans'' are structural  elements  not  observed  in 
spiral galaxies. They may be due to magnetic fields pulled out from  the 
central star-forming region by gas outflows.  Evidence  for  radial  gas 
flows in NGC~4449 was indeed found by Martin (\cite{martin98}, \cite{martin99}).  However,  in 
case of an initially random magnetic field (e.g. injected by supernovae) 
being stretched by gas flows, the ``fans''  would  contain  interspersed 
magnetic lines directed  towards  and  outwards  from  the  star-forming 
complex, yielding no significant Faraday  rotation  (see  Sect.  4.3.1). 
Thus if the radial ``fans''  would  result  from  the  gaseous  wind,  a 
large-scale,  coherent  preexisting  magnetic  field  would   still   be 
required, like one resulting from the dynamo process. 

Alternatively, the observed magnetic field structure in NGC~4449 can  be 
qualitatively  explained  by  classical  dynamo-generated   fields.   In 
addition to a toroidal field running  around  the  disk,  the  classical 
dynamo process also generates a  poloidal  field  with  lines  of  force 
forming closed loop-like structures perpendicular to the disk plane  and 
with diameters comparable to the galaxy radius  (Donner  \&  Brandenburg 
\cite{donner}). They are due to a radial field component, B$_{r}$, turning into a 
vertical one, B$_{z}$, close to the centre and in  the  disk  outskirts. 
The conservation of magnetic flux leads to  B$_{z}$  being  always  much 
stronger in the central region than in the outer disk. In  large  spiral 
galaxies the vertical segments of the  poloidal  field  loops  with  the 
strongest B$_{z}$ probably lie at heights $\ge 2$ -- 3~kpc. This is  too 
high to see the vertical magnetic field in synchrotron emission, as  the 
latter has a vertical scale height of about 1 kpc (Hummel et  al. \cite{hummel}) 
due to a limited propagation  range  of  cosmic  ray  electrons.  As  an 
exception NGC~4631  has  a  much  larger  scale  height  and  dominating 
vertical fields in its inner regions (Hummel et al. \cite{hummel}). 

With its bright star-forming disk of about 4~kpc  diameter  NGC~4449  is 
several times smaller  than  normal  spirals.  If  it  had  a  classical 
poloidal dynamo-type magnetic  field,  its  magnetic  lines  would  make 
closed loop-like structures with a vertical size of about 1 -- 1.5~kpc. 
The maximum B$_{z}$ would occur at some hundreds of  parsecs  above  the 
galaxy's plane, well within  the  propagation  range  of  radio-emitting 
electrons, making vertical fields visible in emission. The intense  star 
formation in NGC~4449 and its low gravitational potential may give  rise 
to galactic winds which  may  additionally  enhance  the  generation  of 
vertical magnetic fields (Brandenburg et al. \cite{branden93}). With the inclination 
of NGC~4449 (Table \ref{tab-param}) a strong poloidal field in  the  central  part  of 
NGC~4449, projected to the sky plane,  may give  rise  to  the  observed 
radial magnetic  ``fans''.  A  detailed  MHD  model  of  magnetic  field 
evolution   in   NGC~4449   is   a   subject   of   a   separate   study 
(Otmianowska-Mazur et al., in prep.). We note also that superimposed  on 
the global magnetic field, smaller-scale ($<$ 0.5~kpc or $30\arcsec$  in 
our map) local phenomena (e.g. magnetized shells or giant magnetic loops 
caused by Parker instabilities, Klein et al. \cite{klein96}) may  be  present,  as 
well. They may explain e.g. local RM reversals, like that  seen  in  the 
eastern ``fan'' at RA$_{2000} \simeq 12^{h} 28^{m} 12^{s}$,  Dec$_{2000} 
\simeq 44\degr 04\arcmin 30\arcsec$. 

Although the dynamo process constitutes some possibility to explain  the 
magnetic field structure in NGC~4449, the question arises how the dynamo 
mechanism can work in this galaxy. Despite some evidence for the  dynamo 
action strong regular magnetic fields are hard to explain  by  classical 
dynamo models which, given the weak signs of rotation of NGC~4449, yield 
growth rates of  the  regular  magnetic  field  at  least  an  order  of 
magnitude smaller than in rapidly rotating spirals (see e.g. Brandenburg 
\& Urpin \cite{branden98}). Estimates  kindly provided by Dr Anvar Shukurov indicate 
that for the rotation speed and dimensions of  NGC~4449  the  classical, 
Coriolis force-driven $\alpha$-effect is  too  weak  for  the  onset  of 
either $\alpha-\omega$ or $\alpha^2$ dynamo (see Ruzmaikin et  al. \cite{ruzmaikin} 
for definitions). Faster field amplification is predicted  by  a  recent 
concept of the dynamo driven by magnetic  buoyancy  and  sheared  Parker 
instabilities (e.g. Moss et al. \cite{moss}). Crude estimates of its efficiency 
by A. Shukurov (priv. comm.) show that the $\alpha^2$ dynamo process  is 
easily excited throughout most of the galaxy's body. However, what  kind 
of structure is generated in  such  conditions  remains  still  an  open 
question and will be a subject  of  separate  analytical  and  numerical 
studies. 

Among other possibilities we  can  mention  e.g.  fast  dynamos  (Parker 
\cite{parker}), interrelations between small-scale velocity  and  magnetic  field 
perturbations caused by specific  instabilities  (Brandenburg  \&  Urpin 
\cite{branden98}) or even magnetic field amplification without  any  $\alpha$-effect 
at all (Blackman \cite{blackman}). As in these concepts ordered rotation  is  still 
needed it is not known how they would work in the complex velocity field 
of NGC~4449. In summary, our  work  provides  arguments  in  support  of 
non-standard magnetic field generation mechanisms, though some  elements 
of its structure may be due to gas outflow processes.  Still  a  lot  of 
theoretical work is needed to understand  how  a  classical  mixture  of 
poloidal and toroidal  fields,  similar  to  that  in  rapidly  rotating 
spirals  can  arise  in  a  slowly  and  chaotically  rotating   object. 
Nevertheless,  it  seems  that  the  existence  of  strong,  dynamically 
important magnetic fields in dwarf irregulars cannot be ignored.

\section{Summary and conclusions} 

We performed a total power and polarization study of the dwarf irregular 
galaxy NGC~4449 at 8.46 and 4.86~GHz using VLA in  its  D-configuration. 
The object rotates slowly and chaotically, thus no  large-scale  regular 
magnetic fields were expected.  To  reach  the  maximum  sensitivity  to 
extended structures we combined our VLA data with the Effelsberg ones at 
10.55~GHz and 4.85~GHz,  respectively.  Despite  the  slow  and  chaotic 
rotation of NGC~4449, unfavourable for dynamo-induced  magnetic  fields, 
we found it to possess strong regular, galaxy-scale fields. 

\bigskip
The following results were obtained: 

\begin{itemize}

\item[-] NGC~4449 shows a large, partly polarized  halo  extending  from 
      its main plane up to 3.5~kpc, more than the isophotal  major  axis 
      radius at 25$^{m}/(\sq\arcsec)$. 

\item[-] The radio-brightest peaks coincide with  strongly  star-forming 
      regions. These regions show increased  thermal  fractions  (up  to 
      80\%), however the nonthermal emission is enhanced there as well. 

\item[-] The galaxy possesses regular magnetic fields reaching locally 6 
      -- 8~$\mu$G,  comparable  to  those  in  rapidly  rotating  spiral 
      galaxies. 
\item[-] NGC~4449 shows  large  domains  of  non-zero  Faraday  rotation 
      measures indicating a genuine galaxy-scale regular magnetic  field 
      rather than random anisotropic ones  with  frequent  reversals  of 
      their direction.  

\item[-] The magnetic field structure consists of  two  basic  elements: 
      radial ``fans'' stretching  away  from  the  central  star-forming 
      complex and a magnetic ring at the radius of  about  2.2~kpc.  The 
      magnetic field in  the  ring  shows  clear  characteristics  of  a 
      magnetic spiral with a  substantial  radial  component  signifying 
      dynamo action. 

\item[-] Both the radial ``fans'' and the polarized ring  can still  be 
      explained in terms of a combination of sky-projected poloidal  and 
      toroidal dynamo-generated fields, taking into account the  smaller 
      size   of   NGC~4449   compared   to   normal   massive   spirals. 
      Alternatively, magnetic ``fans'' could result from the gas outflow 
      from the  central  star-forming  complex.  Even  in  this  case  a 
      large-scale coherence of the magnetic field subject to  stretching 
      by outflows is required. Thus,  some  kind  of  dynamo  action  is 
      needed, with a preference of non-standard  (e.g.  buoyancy-driven) 
      dynamos. Whether  and  how  this  process  can  produce  classical 
      dynamo-like magnetic fields in  a  complex  and  chaotic  velocity 
      field of NGC~4449 remains yet unknown. 

\end{itemize} 

The detection of regular magnetic fields in spiral galaxies is important 
for understanding processes like turbulence, turbulent diffusion and the 
magnetic field generation in astrophysical  plasmas;  this  is  also  of 
importance for plasma physics in general. It demonstrated that even in a 
highly turbulent medium large-scale regular fields can persist and  grow 
quite efficiently. This has already boosted the  development  of  dynamo 
theories applicable not only to a variety of astrophysical objects  from 
planets to clusters of galaxies but also to laboratory plasmas.  On  the 
other hand, there was a widespread  prejudice  that  all  the  mentioned 
concepts are  restricted  solely  to  rapidly  rotating  plasma  bodies. 
Against these expectations we show that strong regular fields  can  also 
arise in slowly and chaotically rotating systems. Their  dynamical  role 
in dwarf irregulars, especially in processes of star-formation triggered 
by magnetic instabilities, filament formation and  confinement  or  even 
accelerating galactic winds via cosmic ray pressure exerted on MHD waves 
(Breitschwerdt et al. \cite{breit}), cannot be further  neglected.  NGC~4449  is 
the irregular galaxy with the best studied magnetic  field  so  far.  We 
believe that further  progress  needs  more  detailed  models  for  such 
objects. Further detailed observations of the radio  polarization  of  a 
larger number of irregulars with various  morphological  characteristics 
are also required. 

\begin{acknowledgements} 
The Authors wish to express their  thanks  to  Dr  Dominik  Bomans  from 
Astronomisches Institut der Ruhr-Universit\"at Bochum for  providing  us 
with his H$\alpha$~map  in  a  numerical  format.  We  are  grateful  to 
numerous colleagues from the Max-Planck-Institut  f\"ur  Radioastronomie 
(MPIfR) in Bonn for their valuable discussions during this work. We want 
to express our profound gratitude to Dr Elly M. Berkhuijsen  from  MPIfR 
for her critical reading of  the  manuscript  and  precious  suggestions 
concerning its improvement. M.U. and K.Ch. are indebted to Professor  R. 
Wielebinski (MPIfR) for the invitations to stay at this institute  where 
substantial parts of this work were done. One of us (K.Ch.) is  indebted 
to Professor Miller Goss from NRAO for his invitation  to  Soccorro  and 
his assistance in some parts of data reduction. We are also grateful  to 
colleagues  from  the  Astronomical  Observatory  of  the   Jagiellonian 
University in Krak\'ow for their comments. This work was supported by  a 
grant from the Polish Research Committee (KBN), grant no. 962/P03/97/12. 
Large parts of computations were made using the HP715 workstation at the 
Astronomical Observatory in Krak\'ow, partly sponsored by the ESO  C\&EE 
grant A-01-116 and on the Convex-SPP machine at  the  Academic  Computer 
Centre "Cyfronet"  in  Krak\'ow  (grant  no.  KBN/C3840/UJ/011/1996  and 
KBN/SPP/UJ/011/1996). 
\end{acknowledgements}

\end{document}